\documentclass[preprint,prd,aps,showpacs,showkeys,nofootinbib]{revtex4-1}
\usepackage{amssymb}
\usepackage{amsmath}
\usepackage{graphicx}
\usepackage{subfigure}
\usepackage{float}

\begin{document}

%\preprint{hep-ph/0412147}

\title{Inflection point in running kinetic term inflation}

\author{Tie-Jun Gao$^1$}
\email{tjgao@xidian.edu.cn}, \author{Wu-Tao Xu$^2$}\email{xwutao@itp.ac.cn}, \author{Xiu-Yi Yang$^3$}\email{yxyruxi@163.com}

\affiliation{$^1$School of Physics and Optoelectronic Engineering, Xidian University, Xi'an 710071, China\\
$^2$CAS Key Laboratory of Theoretical Physics, Institute of Theoretical Physics, Chinese Academy of Sciences, Beijing, 100190, China\\
$^3$School of science, University of Science and Technology Liaoning, Anshan 114051, China}

\begin{abstract}
In this work, we consider inflection point inflation in the framework of running kinetic term inflation in supergravity.
We study the inflationary dynamics and show that the predicted value of the scalar spectral index and tensor-to-scalar ratio can lie within the $1\sigma$ confidence region allowed by the result of Planck 2015.% which is better then the original chaotic inflation model.

\end{abstract}

 \pacs{}

\maketitle

\section{Introduction\label{sec1}}

Inflation was proposed to address a series of questions, such as horizon problem, flatness problem, monopole problems and so on.  Moreover  cosmological inflation is now getting established by all observational data such as the WMAP~\cite{ref1} and Planck space missions~\cite{ref2}. Last year, the full-mission Planck observations of the cosmic microwave background radiation anisotropies constrain the spectral index of curvature perturbations and the tensor-to-scalar ratio to be $n_s=0.9655\pm0.0062$ and $r_{0.002}<0.10$ at $95\%$ confidence level~\cite{ref2}, respectively, which will narrow down some inflation models.

Among various ways of inflation models building, an interesting framework is embed the inflationary models into a more fundamental theory of quantum gravity, such as supergravity. However, most of the old inflation models constructed in supergravity~\cite{ref3,ref4,ref5} suffer from the $\eta$ problem~\cite{ref6}: The F-term potential is proportional to an exponent factor $e^{|\phi|^2}$, which gives a contribution to one of the slow-roll parameter $\eta$, and breaks the slow-roll condition. This problem can be solved by several methods  ~\cite{ref7,ref8,ref9,ref10}. One of the methods is to add a shift-symmetric $\phi\rightarrow \phi+C$ to the K\"{a}hler potential $K(\phi-\phi^{\dag})$ \cite{ref11,ref12,ref13,ref14,ref15}. So the inflaton will not appear in the K\"{a}hler potential, then the exponent factor disappear and guarantee the flatness of the potential. In this case, one can get the normal chaotic inflation with a power-law potential $V(\phi)\propto\phi^n$.

In Ref.\cite{ref16,ref17,ref170}, the author extend the Nambu-Goldstone-like shift symmetry of chiral superfield $\phi$ to a  generalized form $\phi^n\rightarrow \phi^n+C$, then the K\"{a}hler potential is a function of $(\phi^n-\phi^{n\dag})$. In order to get a reasonable potential, one should consider  small shift symmetry breaking terms  $\kappa|\phi|^2$ in the  K\"{a}hler potential and $\lambda\phi^mX$ in the  superpotential respectively, with $\kappa,\lambda\ll1$, and $X$ is an extra superfield. Then one can get a fraction-power scalar potential $V(\phi)\propto\phi^{m/n}$,  the scalar spectral index $n_s$ and tensor-to-scalar ratio $r$ can be approximate as $n_s=1-(1+\frac{m}{n})\frac{1}{N}$, $r=\frac{8m}{n}\frac{1}{N}$\cite{ref17}. Since the coefficient of the kinetic term in this model is field-dependent: $(\kappa+n^2|\phi|^{(2n-2)})\partial_\mu\phi\partial^{\mu}\phi^{\dag}$, it is so called  running kinetic term inflation.

 The inflection point inflation in the framework of MSSM was for the first time realized by the gauge invariant flat directions $udd$ or $LLe$ \cite{ref171}, and developed in \cite{ref172,ref173,ref174,ref175}. In Ref.\cite{ref176} we constructed an inflection point inflationary model with a single chiral superfield, which are consistent with the Planck 2015 result. And after inflation, the model has a non-SUSY de-Sitter vacuum responsible for  the recent accelerated expansion of the Universe.

 In this paper, we will consider the possibility to construct an inflection point inflationary model with a generalized shift symmetry. We first calculate the general form of the scalar potential with polynomial superpotential in the framework of running kinetic term inflation, then focus on a polynomial superpotential of the form  $W= X(\lambda_p\phi^p+\lambda_qe^{i\theta}\phi^q)$ and  setup the inflection point inflationary model. We study the inflationary dynamics and show that the predicted value of the scalar spectral and tensor-to-ratio  are consistent with the result of Planck 2015 at the $95\%$ confidence level. After the end of inflation, the inflaton starts to oscillate around the origin, and then decays into SM particles to reheat the Universe.

The outline of this paper is as follows: In the next section, we give the general form of the scalar potential with polynomial superpotential. In Section 3, we focus on the polynomial superpotential having two terms and setup the inflection point inflationary model. In Section 4, we discuss  the inflationary  dynamics  of the model.%, and the reheating process after inflation.% Final section is devoted to summary.

\section{Polynomial superpotential with running kinetic term \label{sec2}}

In this section, we will calculate the scalar potential of polynomial superpotential in the framework of running kinetic term  inflation model in supergravity. In such model, the K\"{a}hler potential satisfy the generalized shift symmetry of chiral superfield $\phi$ :
\begin{eqnarray}
&&\phi^n\rightarrow \phi^n+C,
\label{kp}
\end{eqnarray}
thus it must be a function of  $i\chi\equiv(\phi^n-\phi^{n\dag})$:
\begin{eqnarray}
&&K=\sum _{l=1}\frac{c_l}{l}(\phi^n-\phi^{n \dag})^l=ic_1(\phi^n-\phi^{n \dag})-\frac{1}{2}(\phi^n-\phi^{n \dag})^2+\cdots.
\label{kp}
\end{eqnarray}
with coefficient $c_l$ is a number  of the order of unity. During inflation, we can prove that $\chi$ is stabilized at the minimum, so
the terms with $l\geq3$ does not change the form of the kinetic term significantly, thus we neglect them\cite{ref16}.  And for $l\leq2$ , $\chi$ is stabilized at $\chi_{min}\simeq c_1$ during inflation.

 In order to give a successful inflation, a small  shift symmetry breaking term are necessary for the K\"{a}hler potential:
  \begin{eqnarray}
&&\Delta K=\kappa|\phi|^2,
\label{kp}
\end{eqnarray}
with $0<\kappa\ll1$ .

Different from Ref.\cite{ref17}, we  assumed the superpotential to be a more  general form \cite{ref18,ref19,ref20,ref21,ref15}
  \begin{eqnarray}
&&W= X(\sum_{m=0}\lambda_m\phi^m)+W_0.
\label{kp}
\end{eqnarray}
Here $\lambda_m\ll1$, and  $X$ is an extra chiral superfield, which is assumed to be stabilized at $X=0$ by higher order terms in the  K\"{a}hler potential. And $|W_0|\simeq m_{3/2}$  with $m_{3/2}$ is the gravitino mass, which is assumed much smaller than the Hubble parameter during inflation, so it can be dropped in the discussion of the inflationary dynamics.

So the total K\"{a}hler potential are
\begin{eqnarray}
&&K=\kappa|\phi|^2+c_1(\phi^n-\phi^{n \dag})-\frac{1}{2}(\phi^n-\phi^{n \dag})^2+|X|^2\cdots.
\label{kp}
\end{eqnarray}
where the dots denote the higher order stabilized term of $X$.

The effective  Lagrangian for the scalar $\phi$ is determined by  superpotential  $W$ as well as  K\"{a}hler potential, which is given by
\begin{eqnarray}
&&L=-K^{\phi\phi^{\dag}}\partial_\mu\phi\partial^{\mu}\phi^{\dag}-V\nonumber\\
&&\hspace{0.4cm}=(\kappa+n^2|\phi|^{(2n-2)})\partial_\mu\phi\partial^{\mu}\phi^{\dag}-V,
\label{infp}
\end{eqnarray}
 and the scalar potential of the inflaton can be calculated by
\begin{eqnarray}
&&V=e^K\Big[D_{\phi}W(K^{-1})^{\phi\phi^{\dag}}(D_{\phi}W)^*-3|W|^2\Big]\nonumber\\
&&\hspace{0.4cm}=e^{\kappa|\phi|^2+c_1(\phi^n-\phi^{n \dag})-\frac{1}{2}(\phi^n-\phi^{n \dag})^2}(\sum_m\lambda_m|\phi|^{m})^2,
\label{infp}
\end{eqnarray}
where
\begin{eqnarray}
&&D_{\phi}W=\partial_{\phi}W+(\partial_{\phi}K)W,
\end{eqnarray}
and $(K^{-1})^{\phi\phi^{\dag}}$ is the inverse of the K\"{a}hler metric
\begin{eqnarray}
&&K^{\phi\phi^{\dag}}=\frac{\partial^2K}{\partial\phi\partial\phi^{\dag}}
\end{eqnarray}

For $(\kappa/n^2)^{1/(2n-2)}\ll|\phi|\ll\kappa^{-1/2}$, the $\kappa$-term can be neglected, so  we can define $\hat{\phi}\equiv\phi^n$, which means that $\hat{\phi}$ is invariant under a Nambu-Goldstone like shift symmetry, and then the  Lagrangian (6) can be approximated by
\begin{eqnarray}
&&L=\partial_\mu\hat{\phi}\partial^{\mu}\hat{\phi}^{\dag}-e^{-\frac{|c_1|^2}{2}}(\sum_m\lambda_m|\hat{\phi}|^{m/n})^2\nonumber\\
&&\hspace{0.4cm}
=\partial_\mu\hat{\phi}\partial^{\mu}\hat{\phi}^{\dag}-(\sum_m\hat{\lambda}_m|\hat{\phi}|^{m/n})^2,
\label{infp}
\end{eqnarray}
where we have used $i\chi\equiv(\phi^n-\phi^{n\dag})=ic_1$,   $\hat{\lambda}_m\equiv e^{-\frac{|c_1|^2}{4}}\lambda_m$.

Because of the shift symmetry, the  real component of $\hat{\phi}$ does not appear in the K\"{a}hler
potential, so the potential is considerably flat along the real direction and thus  becomes an
inflaton candidate. the imaginary component acquires a mass heavier than the Hubble scale and is stabilized at the minimum  during inflation. Therefore, we can set $\hat{\phi}\equiv(\varphi+i\chi)/\sqrt{2}=\varphi/\sqrt{2}+ic_1/2$ and obtain the scalar potential of $\varphi$, which has the form
\begin{eqnarray}
&&V=\Big[\sum_{m=0}\hat{\lambda}_m\Big(\frac{\varphi}{\sqrt{2}}\Big)^{\frac{m}{n}}\Big]^2.
\label{infp}
\end{eqnarray}

\section{Setup of the inflection point inflation \label{sec2}}

In order to construct an inflection point inflationary model, we will focus on a polynomial superpotential having two terms
\begin{eqnarray}
&&W= X(\lambda_p\phi^p+\lambda_qe^{i\theta}\phi^q),
\label{kp}
\end{eqnarray}
with the powers $q>p\geq1$ are integer, and the coefficient $\lambda_p, \lambda_q$ is positive without loss of generality,  $\theta$ is the phase of the second term.

Using Eq.(11), the scalar potential corresponding to the superpotential (12) can be written by
\begin{eqnarray}
&&V=\Big[\hat{\lambda}_p\Big(\frac{\varphi}{\sqrt{2}}\Big)^{\frac{p}{n}}+\hat{\lambda}_qe^{i\theta}\Big(\frac{\varphi}{\sqrt{2}}\Big)^{\frac{q}{n}}\Big]^2\nonumber\\
&&\hspace{0.4cm}=\hat{\lambda}_p^2\Big(\frac{\varphi}{\sqrt{2}}\Big)^{\frac{2p}{n}}\Big[1+2\xi\cos\theta\Big(\frac{\varphi}{\sqrt{2}}\Big)^{\frac{q-p}{n}}
+\xi^2\Big(\frac{\varphi}{\sqrt{2}}\Big)^{\frac{2(q-p)}{n}}\Big],
\label{infp}
\end{eqnarray}
where  $\hat{\lambda}_{p,q}\equiv e^{-\frac{|c_1|^2}{4}}\lambda_{p,q}$
 and $\xi=|\hat{\lambda}_q/\hat{\lambda}_p|=|\lambda_q/\lambda_p|$.  The  curve of the  potential is shown in Fig.1.

%%%%%%%%%%%%%%%%%%%%%%%%%%%%%%%%%%%%%%%%%%%%%%%%%%%%%%%%%%%%%%%%%%%%%%%%%%%%%%%%%%%%5
\begin{figure}[H]\small

  \centering
   \includegraphics[width=4in]{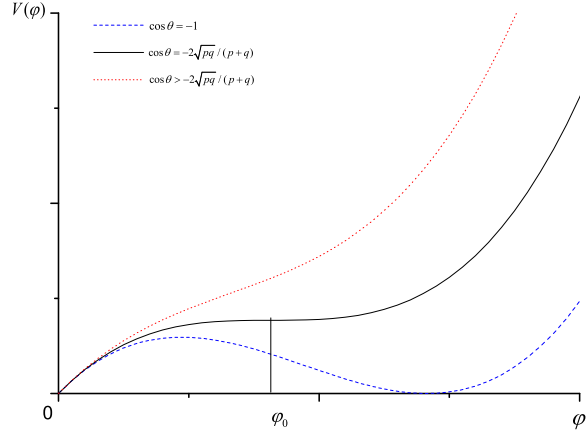}
     \caption{The inflation potential $V(\varphi)$ with $\cos\theta=-1$(dash line), $\cos\theta=-2\sqrt{pq}/(p+q)$ (solid line ), $\cos\theta>-2\sqrt{pq}/(p+q)$ (dot line), respectively.}
    \label{fig-sp}
\end{figure}
%%%%%%%%%%%%%%%%%%%%%%%%%%%%%%%%%%%%%%%%%%%%%%%%%%%%%%%%%%%%%%%%%%%%%%%%%%%%%%%%%%%%%%%%

We could see that if $\cos\theta=-1$, there are one minimum at  $\varphi=\sqrt{2}(\frac{p}{\xi q})^{\frac{n}{q-p}}$ and one minimum or zero value at $\varphi=0$ respectively, which is shown in Fig.1(dash line).
As $\cos\theta$ increase, the minimum at $\varphi=\sqrt{2}(\frac{p}{\xi q})^{\frac{n}{q-p}}$   is uplift. In this case, the initial value of the inflation  field should below the local maximum, otherwise the inflaton will be trapped in the false vacuum.

An interest case is that if the parameters $\theta$ satisfies the relation
\begin{eqnarray}
\cos\theta=-\frac{2\sqrt{pq}}{p+q}
\end{eqnarray}
the local maximum and the second minimum will be the same at  point $\varphi_0=\sqrt{2}\Big(\frac{1}{\xi}\sqrt{\frac{p}{q}}\Big)^{\frac{n}{q-p}}$, and the false vacuum disappears(solid line), this point is called inflection point. At this point, the inflation potential is
\begin{eqnarray}
V(\varphi_0)=\hat{\lambda}_p^2 \left(\frac{p}{q}+\frac{4 p}{p+q}+1\right) \left(\frac{p}{\xi ^2 q}\right)^{\frac{p}{q-p}},
\end{eqnarray}
and both the first and second derivatives of $V$ vanish at $\varphi_0$. It's worth mentioning that this potential is independent of $n$, so for different $n$, the potential at inflection point are almost the same.  And we will see shortly that since there is a flat plateau at the inflection point  $\varphi_0$, the predicted spectral index as well as the tensor-to-scalar ratio can lie within the $1\sigma$ confident region allowed by Planck 2015.

When $\cos\theta>-\frac{2\sqrt{pq}}{p+q}$, then the flat plateau disappear, and  power law  inflation model will be reproduced(dot line ).

It is worth mentioning that the potential near the origin is an odd function for some choice of parameters, which will cause undesirable behavior after inflation. It doesn't  matter since when the scalar field $\varphi<\sqrt{2}(\kappa/n^2)^{n/(2n-2)}$, the first term of the kinetic term $\kappa$ in (6) become important, so we couldn't drop it, and the scalar $\phi$ becomes the dynamical variable. Then  the potential in this region is not as in Fig.1, which will have the form $(\sum_m\lambda_m|\phi|^{m})^2$. And as $\phi$  decrease, the soft SUSY breaking mass term $m_{\phi}|\phi|^2$ will dominates the potential\cite{ref17}, so the scalar will oscillate  around the origin and reheat the Universe.

In this paper we focus on the inflation potential with inflection point. Since the parameter $\theta$ satisfy the relation(14), so for a given $n, p$ and $q$, there are only two free parameters $\hat{\lambda}_p$ and $\xi$. The inflation potential~(13) becomes
\begin{eqnarray}
&&V(\varphi)=\hat{\lambda}_p^2\Big(\frac{\varphi}{\sqrt{2}}\Big)^{\frac{2p}{n}}\Big[1-4\xi\frac{\sqrt{pq}}{p+q}\Big(\frac{\varphi}{\sqrt{2}}\Big)^{\frac{q-p}{n}}
+\xi^2\Big(\frac{\varphi}{\sqrt{2}}\Big)^{\frac{2(q-p)}{n}}\Big].
\label{infp}
\end{eqnarray}

\section{slow-roll inflation  \label{sec2}}

The slow-roll parameters $\epsilon$ and $\eta$ are defined  by
\begin{eqnarray}
&&\epsilon \equiv \frac{1}{2}\left(\frac{V'(\varphi)}{V(\varphi)}\right)^2 ,\nonumber\\
&& \eta \equiv \frac{V''(\varphi)}{V(\varphi)}.
\end{eqnarray}
To first order in the slow-roll approximation, the scalar spectral index $n_s$ and tensor-to-scalar ratio $r$ are expressed in terms of the slow-roll parameters:
\begin{eqnarray}
&&n_s \simeq 1-6\epsilon+2\eta ,\nonumber\\
&&r \simeq 16\epsilon.
\end{eqnarray}
The number of $e$-folding during inflation can be written  by
\begin{eqnarray}
&&N=\int^{\varphi_i}_{\varphi_f}\frac{V}{V'}d\varphi ,
\end{eqnarray}
and the field value at the end of inflation $\varphi_f$ is determined by Max$\{\epsilon(\varphi_f),\eta(\varphi_f)\}=1$.

The parameter $\hat{\lambda}_p$ is constrained by the amplitude of curvature perturbations:
\begin{eqnarray}
&&\Delta_R^2=\frac{V}{24\pi^2\epsilon}.
\end{eqnarray}
The maximum likelihood value from the Planck 2015 data is $\Delta_R^2(k_0)=2.19\times10^{-9}$, so for a  given $n, p$ and $q$, we can get the relation between $\hat{\lambda}_p$ and $\xi$. The corresponding figures are show in Fig.2,3,4 and 5 for  different parameters  $n, p$ and $q$ and the e-folding number $N=60$ or $50$ respectively.

%%%%%%%%%%%%%%%%%%%%%%%%%%%%%%%%%%%%%%%%%%%%%%%%%%%%%%%%%%%%%%%%%%%%%%%%%%%%%%%%%%%%5
\begin{figure}\small

  \centering
   \includegraphics[width=4in]{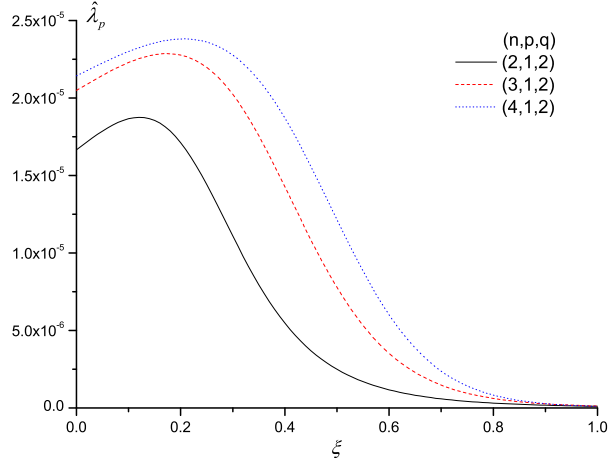}
     \caption{(color online)The value of $\hat{\lambda}_p$ and $\xi$  for reproducing the Planck observed amplitude of curvature perturbations. With the e-folding number $N=60$, and $p, q$ are set to be $(1, 2)$. The black solid line, red dash line and blue dot line  corresponds to the numerical result with the parameters $n=2,3$ and $4$ respectively.}
    \label{Ln,1,2}
\end{figure}
%%%%%%%%%%%%%%%%%%%%%%%%%%%%%%%%%%%%%%%%%%%%%%%%%%%%%%%%%%%%%%%%%%%%%%%%%%%%%%%%%%%%%%%%
%%%%%%%%%%%%%%%%%%%%%%%%%%%%%%%%%%%%%%%%%%%%%%%%%%%%%%%%%%%%%%%%%%%%%%%%%%%%%%%%%%%%5
\begin{figure}\small

  \centering
   \includegraphics[width=4in]{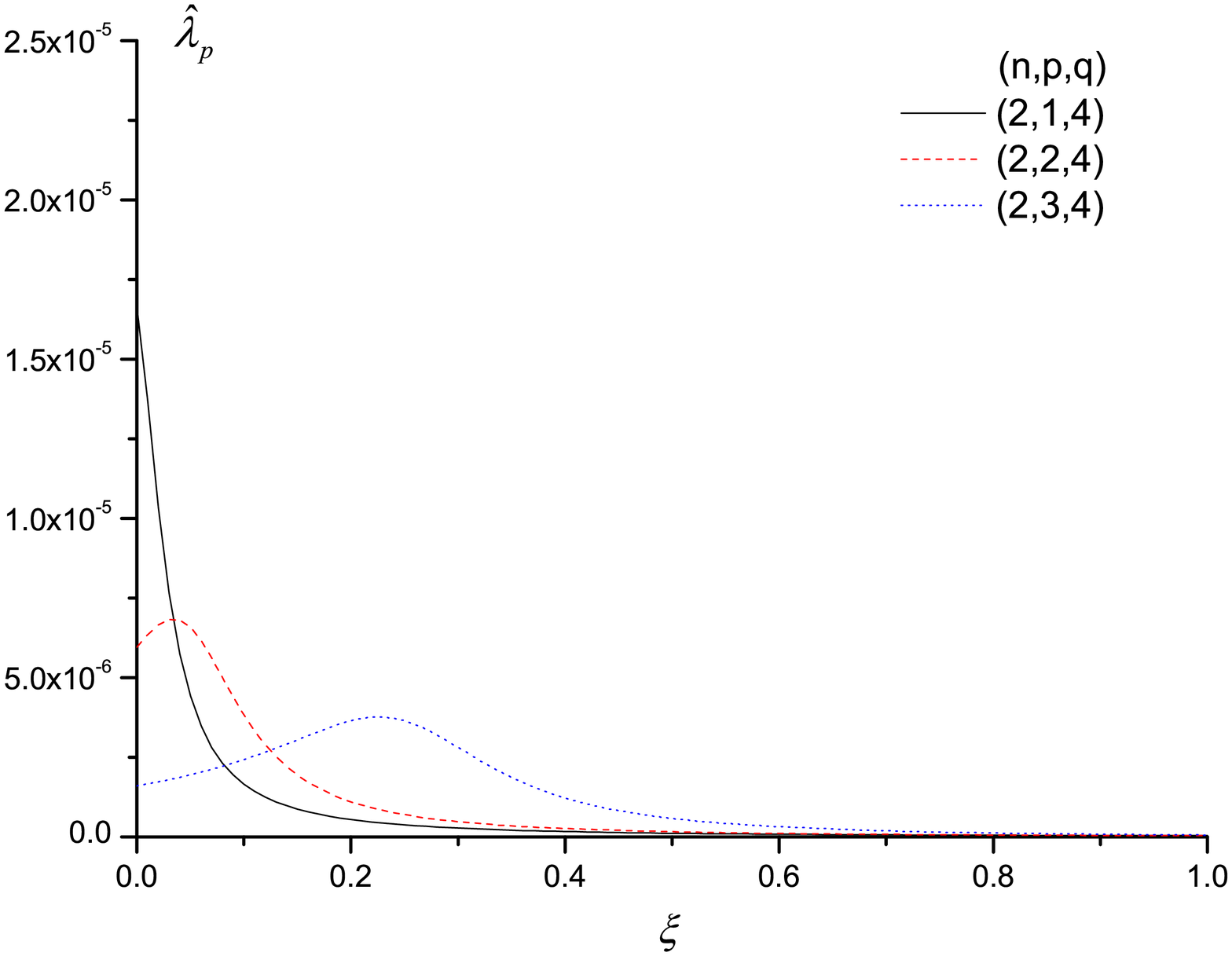}
     \caption{(color online)The value of $\hat{\lambda}_p$ and $\xi$  for reproducing the Planck observed amplitude of curvature perturbations. With the e-folding number $N=60$, and $n, q$ are set to be $(2,4)$. The black solid line, red dash line and dlue dot line  corresponds to the numerical result with the parameters $p=1,2$ and $3$ respectively.}
    \label{L2,p,4}
\end{figure}
%%%%%%%%%%%%%%%%%%%%%%%%%%%%%%%%%%%%%%%%%%%%%%%%%%%%%%%%%%%%%%%%%%%%%%%%%%%%%%%%%%%%%%%%
%%%%%%%%%%%%%%%%%%%%%%%%%%%%%%%%%%%%%%%%%%%%%%%%%%%%%%%%%%%%%%%%%%%%%%%%%%%%%%%%%%%%5
\begin{figure}\small

  \centering
   \includegraphics[width=4in]{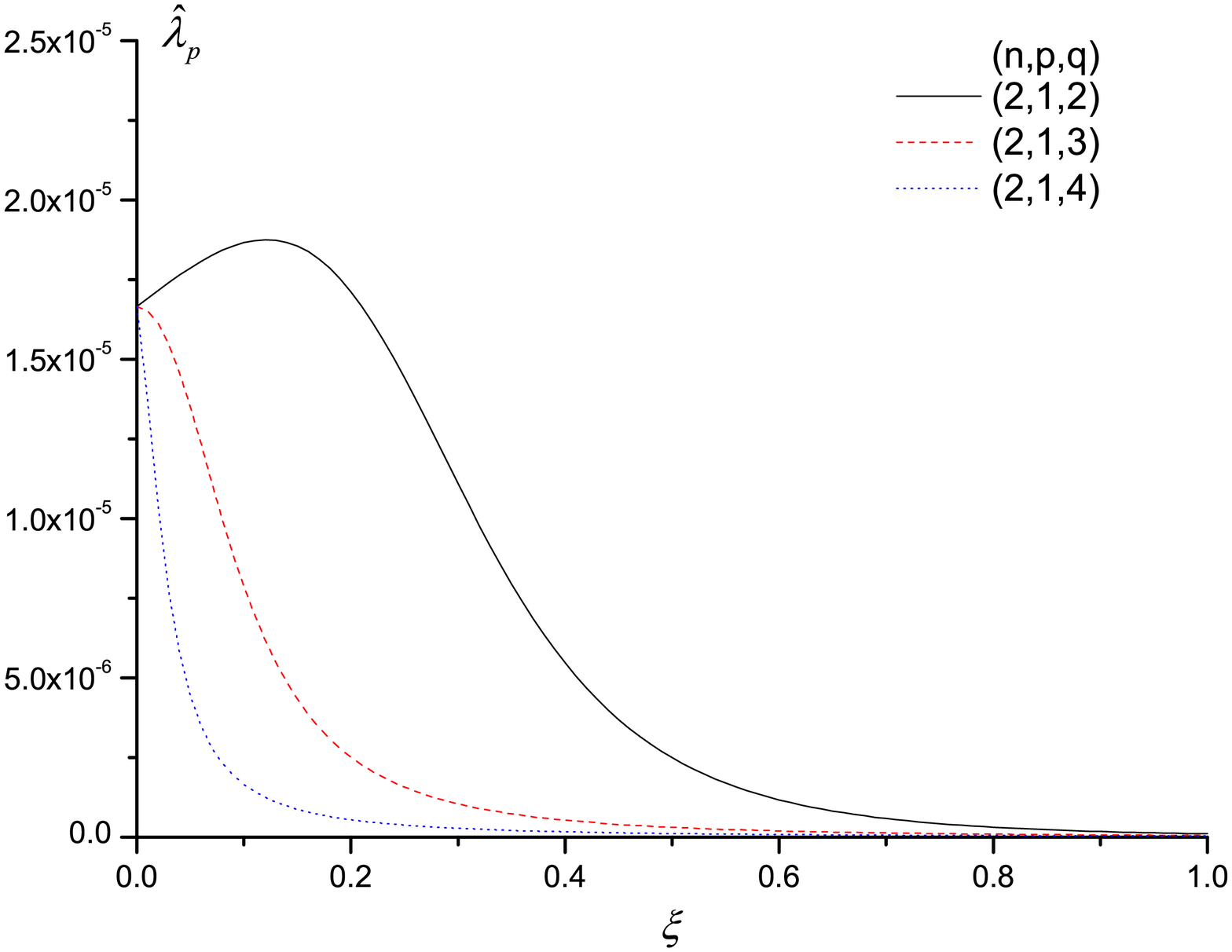}
     \caption{(color online)The value of $\hat{\lambda}_p$ and $\xi$  for reproducing the Planck observed amplitude of curvature perturbations. With the e-folding number $N=60$, and $n,p$ are set to be $(2,1)$. The black solid line, red dash line and blue dot line  corresponds to the numerical result with the parameters $q=2,3$ and $4$ respectively.}
    \label{L2,1,q}
\end{figure}
%%%%%%%%%%%%%%%%%%%%%%%%%%%%%%%%%%%%%%%%%%%%%%%%%%%%%%%%%%%%%%%%%%%%%%%%%%%%%%%%%%%%%%%%
%%%%%%%%%%%%%%%%%%%%%%%%%%%%%%%%%%%%%%%%%%%%%%%%%%%%%%%%%%%%%%%%%%%%%%%%%%%%%%%%%%%%5
\begin{figure}\small

  \centering
   \includegraphics[width=4in]{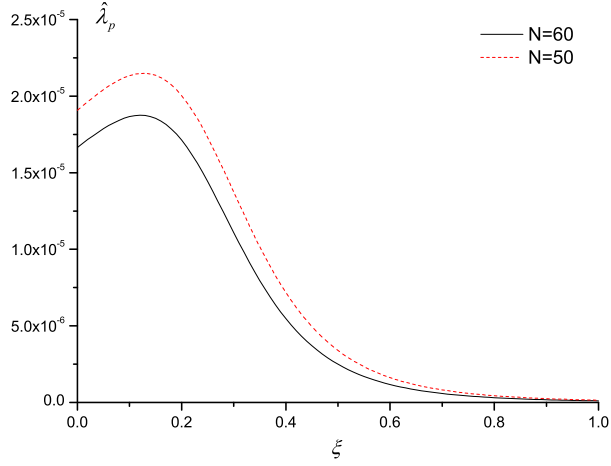}
     \caption{(color online)The value of $\hat{\lambda}_p$ and $\xi$  for reproducing the Planck observed amplitude of curvature perturbations. And $n,p,q$ are set to be $(2,1,2)$. The black solid line and red dash line corresponds to the numerical result with the e-folding number $N=60$ and $N=50$ respectively.}
    \label{LN=50}
\end{figure}
%%%%%%%%%%%%%%%%%%%%%%%%%%%%%%%%%%%%%%%%%%%%%%%%%%%%%%%%%%%%%%%%%%%%%%%%%%%%%%%%%%%%%%%%

%%%%%%%%%%%%%%%%%%%%%%%%%%%%%%%%%%%%%%%%%%%%%%%%%%%%%%%%%%%%%%%%%%%%%%%%%%%%%%%%%%%%5
\begin{figure}\small

  \centering
   \includegraphics[width=4in]{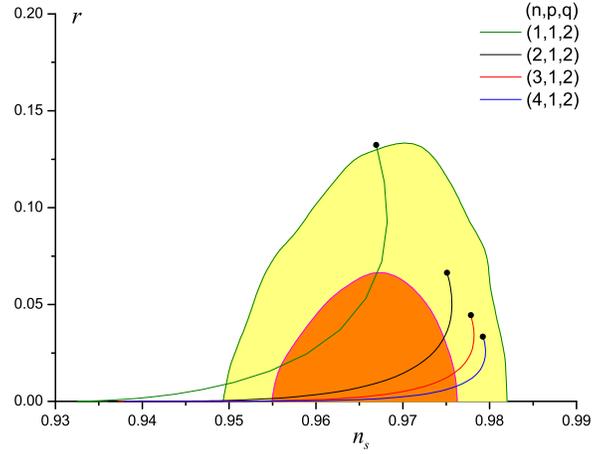}
     \caption{(color online)The $n_s-r$ region predicted by the model with the e-folding number $N=60$. The contours are the marginalized joint 68\% and 95\% confidence level regions for $n_s$ and $r$ at the pivot scale $k_*= 0.002$ Mpc$^{-1}$ from the Planck 2015 TT+lowP data.  The parameters $p,q$ are set to be (1,2), and the power $n$ is taken as $n=1,2,3$ and $4$ from top to bottom. The dots on the line correspond to the case of $\xi=0$. }
    \label{Gn,1,2}
\end{figure}
%%%%%%%%%%%%%%%%%%%%%%%%%%%%%%%%%%%%%%%%%%%%%%%%%%%%%%%%%%%%%%%%%%%%%%%%%%%%%%%%%%%%%%%%
%%%%%%%%%%%%%%%%%%%%%%%%%%%%%%%%%%%%%%%%%%%%%%%%%%%%%%%%%%%%%%%%%%%%%%%%%%%%%%%%%%%%5
\begin{figure}\small

  \centering
   \includegraphics[width=4in]{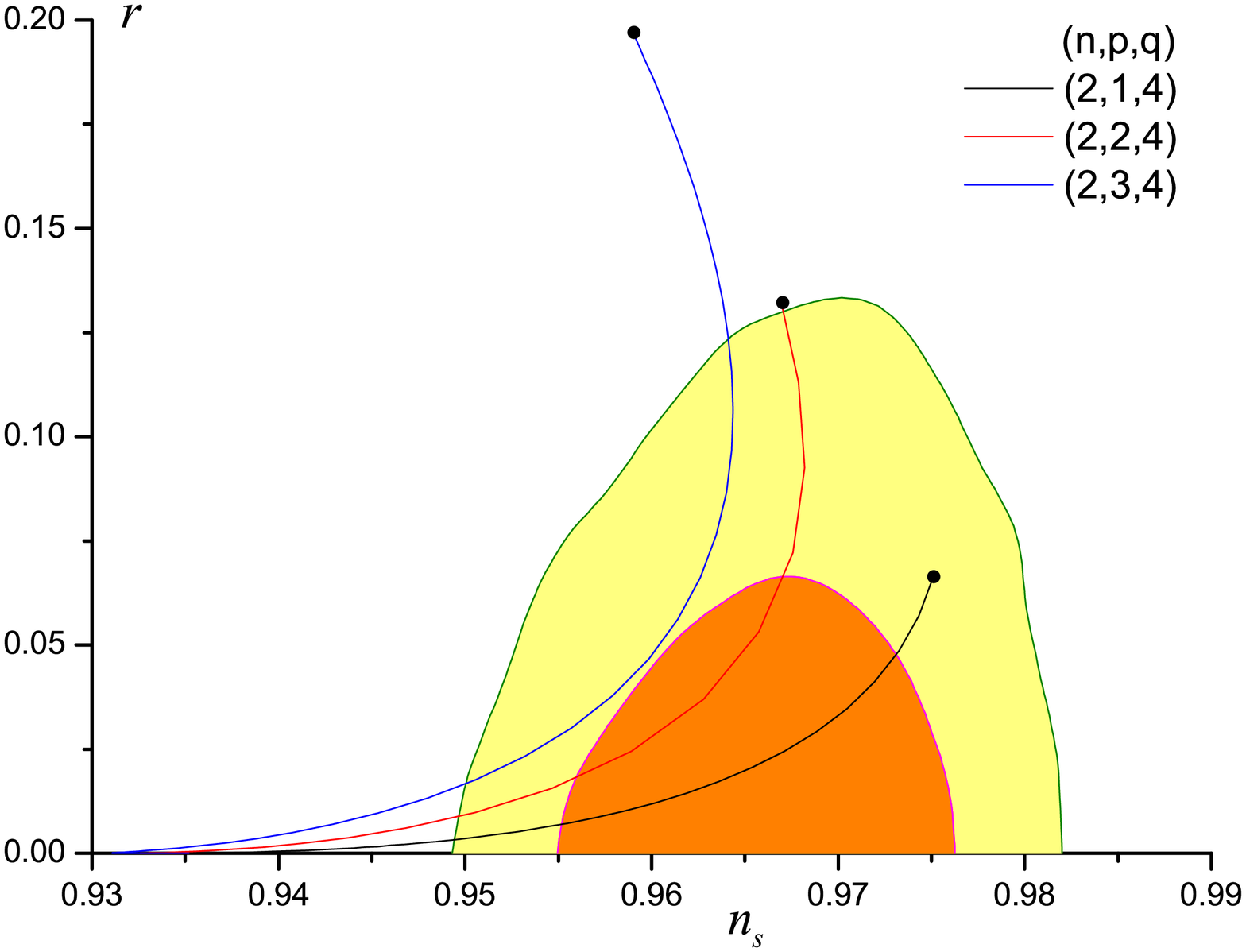}
     \caption{(color online)The $n_s-r$ region predicted by the model with the e-folding number $N=60$. The parameters $n,p$ are set to be (2,4), and the power $p$ is taken as $p=3,2,1$ from top to bottom. The dots on the line correspond to the case of $\xi=0$.}
    \label{2,p,4}
\end{figure}
%%%%%%%%%%%%%%%%%%%%%%%%%%%%%%%%%%%%%%%%%%%%%%%%%%%%%%%%%%%%%%%%%%%%%%%%%%%%%%%%%%%%%%%%
%%%%%%%%%%%%%%%%%%%%%%%%%%%%%%%%%%%%%%%%%%%%%%%%%%%%%%%%%%%%%%%%%%%%%%%%%%%%%%%%%%%%5
\begin{figure}\small

  \centering
   \includegraphics[width=4in]{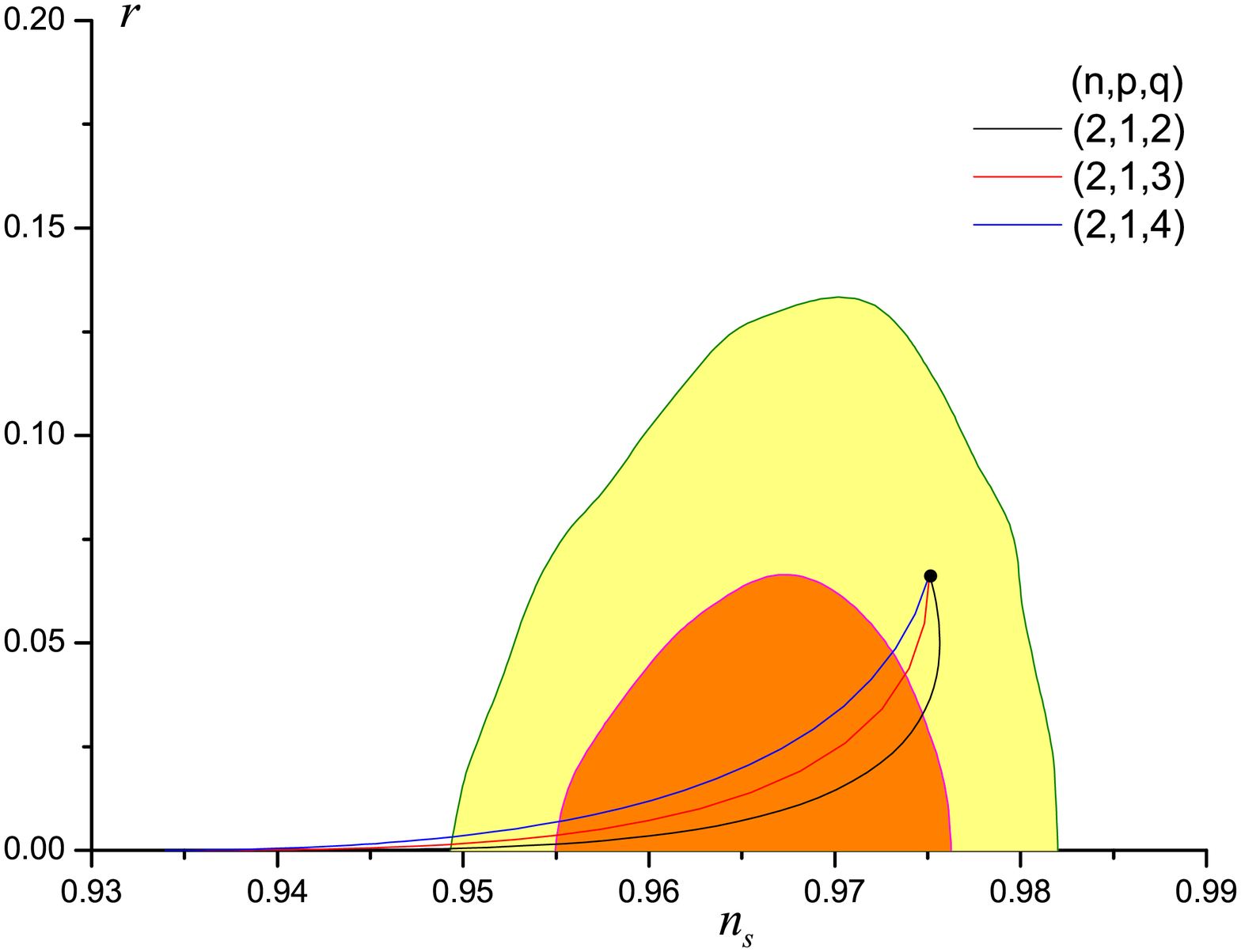}
     \caption{(color online)The $n_s-r$ region predicted by the model with the e-folding number $N=60$. The parameters $n,p$ are set to be (2,1), and the power $q$ is taken as $q=4,3,2$ from top to bottom. The dots on the line correspond to the case of $\xi=0$.}
    \label{G2,1,q}
\end{figure}
%%%%%%%%%%%%%%%%%%%%%%%%%%%%%%%%%%%%%%%%%%%%%%%%%%%%%%%%%%%%%%%%%%%%%%%%%%%%%%%%%%%%%%%%
%%%%%%%%%%%%%%%%%%%%%%%%%%%%%%%%%%%%%%%%%%%%%%%%%%%%%%%%%%%%%%%%%%%%%%%%%%%%%%%%%%%%5
\begin{figure}\small

  \centering
   \includegraphics[width=4in]{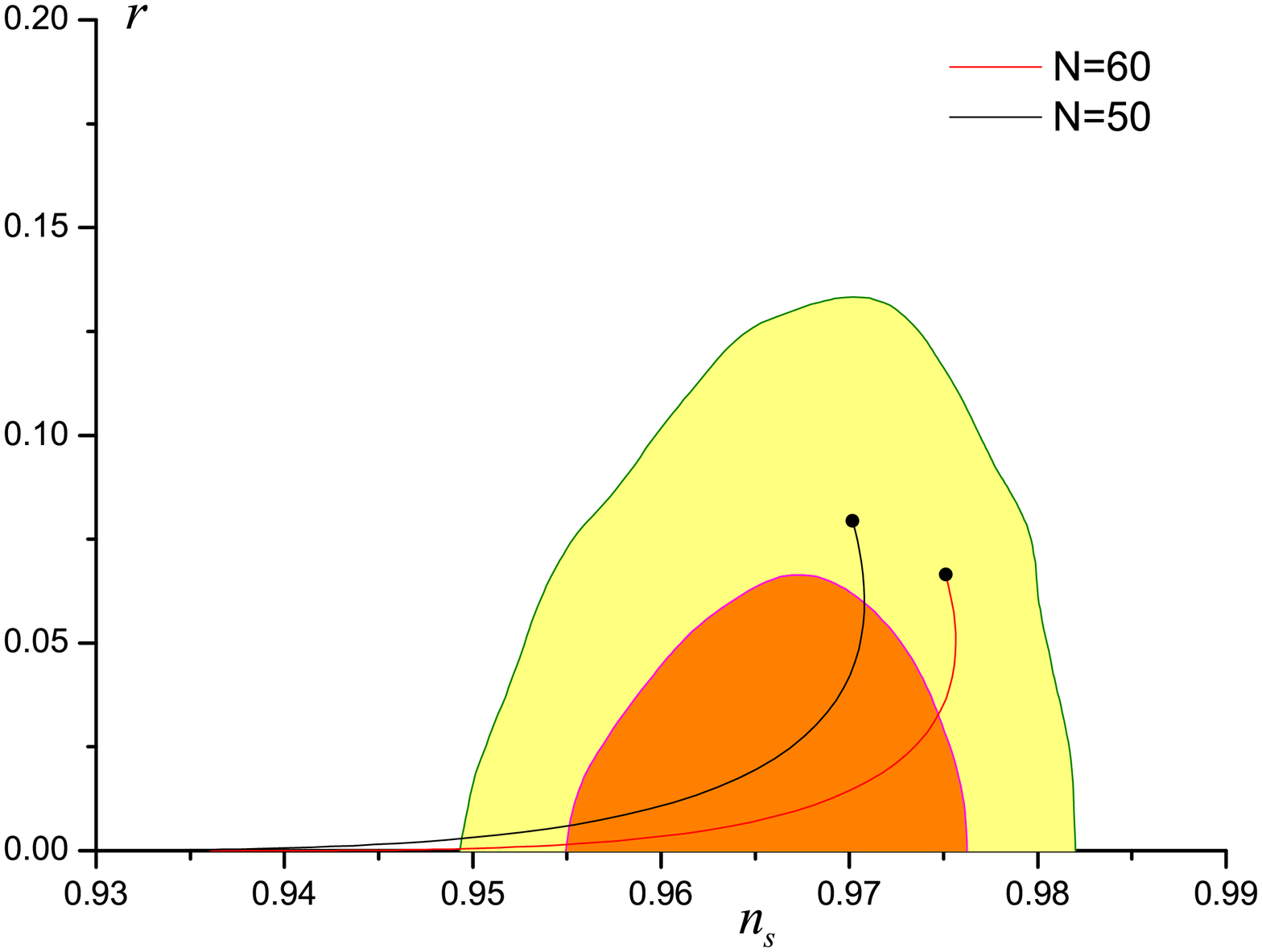}
     \caption{(color online)The $n_s-r$ region predicted by the model. The parameters $n$, $p$, $q$ are set to be (2,1,2). The black line and red line corresponds to the numerical result with the e-folding number $N=60$ and $N=50$ respectively. }
    \label{GN=50}
\end{figure}
%%%%%%%%%%%%%%%%%%%%%%%%%%%%%%%%%%%%%%%%%%%%%%%%%%%%%%%%%%%%%%%%%%%%%%%%%%%%%%%%%%%%%%%%

 Fig.6,7,8 and 9 show the $n_s-r$ region  predicted by the model for different parameters $n$, $p$ and $q$.   The contours are the marginalized joint 68\% and 95\% confidence level regions for $n_s$ and $r$ at the pivot scale $k_*=0.002$ Mpc$^{-1}$ from the Planck 2015 TT+lowP data. We could see that the curves are consistent with the Planck data. The dots on the line correspond to the case of  $\xi=0$,  which is just the case of origin running kinetic term  inflationary model  with potential $V\propto\varphi^{m/n}$\cite{ref17},  and as $\xi$ increase, the $n_s-r$ dot go along the curves to the left.

     From Fig.6, we could see that for a given $p=1$ and $q=2$, as $n$ increase from $1$ to $4$, the curves become lower and lower, and become better agreement with the Planck data.  Notice that $n=1$ is just the case of the original inflection point inflation. In Fig.7,  we assume the parameters $n=2$, $q=4$, the curves from top to bottom are the power $p=3,2,1$ respectively. In Fig.8, the parameters are assumed $n=2, p=1$,  the curves from top to bottom are the power $q=4,3,2$ respectively. We could see that the three curves converge to one point, the point is just the case of $\xi=0$, which is just the prediction of the origin model with the potential $V\propto\varphi^{m/n}$. The two curves in Fig.9 are the case of $N=60$(black line) and $N=50$(red line) respectively. It can be seen from these figures that the predictions are consistent with the Planck 2015 results.

 After the end of inflation, the scalar $\varphi$ will becomes small, and when $\varphi<\sqrt{2}(\kappa/n^2)^{n/(2n-2)}$, the first term $\kappa$ in the kinetic term (6) becomes more important, then the scalar $\phi$ becomes the dynamical variable. In this region the potential will  have the form $(\sum_m\lambda_m|\phi|^{m})^2$. And as the field $\phi$ decrease, the soft SUSY breaking mass term $m_{\phi}|\phi|^2$ will dominates the potential. The inflaton will oscillate around the origin, and then decays into SM particles by introducing the couplings with Higgs doublets such as $W=\lambda_XXH_uH_d$ or $W=\lambda_\phi\phi H_uH_d$, to reheat the Universe. The reheating process are quite similarly as in Ref.\cite{ref17}.

\section{Summary \label{sec5}}

In this work, the inflection point inflation  in the framework of running kinetic term inflation in supergravity has been successfully constructed using the polynomial superpotential having two terms. The inflationary predictions of the model are consistent with the Planck 2015 results. Such predictions in the $n_s-r$ plane are better then the original model with scalar potential $V\propto\phi^{m/n}$. After inflation, the inflaton will oscillate around the origin and then decays into SM particles by introduce some unsuppressed coupling with SM sector, such as Higgs doublets. %And the reheating temperature can be of order $10$GeV, which is consistently with usual cosmological constraints.%

\begin{acknowledgments}
This work was supported by "the Fundamental Research Funds for the Central Universities"  No.XJS16029 and No.JB160507.
\end{acknowledgments}

\end{document}